\documentstyle[eqsecnum,aps,twocolumn,epsfig,ilja1]{revtex}

\begin{document}
\draft
\title{Ring dark solitary waves: experiment versus theory}
\author{A. Dreischuh, D. Neshev}
\address{Sofia University, Department of Quantum Electronics, 5, J.
Bourchier Blvd., BG-1164 Sofia, Bulgaria\\
(Fax.: +3592/9625276, E-mail: ald@phys.uni-sofia.bg)}
\author{G. G. Paulus$^1$, F. Grasbon$^1$, H. Walther$^{1,2}$}
\address{$^1$Max-Planck-Institut f\"ur Quantenoptik, Hans-Kopfermann-Strasse
1, D-85748 Garching, Germany\\
(Fax.: +4989/329050, E-mail: ggp@mpq.mpg.de)}
\address{$^2$Ludwig-Maximilians-Universit\"at, Sektion Physik, Am Coulombwall 1,
D-85747 Garching, Germany\\
(Fax.: +4989/2891 4142, E-mail: Prof.H.Walther@mpq.mpg.de)}
\date{\today}
\maketitle

\begin{abstract}
Theoretical and experimental
results on optical ring dark solitary waves are presented, emphasizing
the interplay between initial dark beam contrast, phase-shift
magnitude, background-beam intensity and saturation of the nonlinearity
are presented. The results are found to confirm qualitatively the
existing analytical theory and are in agreement with the numerical
simulations carried out.

\end{abstract}

\preprint{HEP/123-qed}

\narrowtext

\section{Introduction}
\label{teil0} Mathematically, optical dark spatial solitons (DSSs)
are exact solutions of the one-dimensional nonlinear Schr\"odinger
equation (NLSE) for negative nonlinearity and nonvanishing
boundary conditions \cite{refc1}. Physically, they form on
background beams of finite width as self-supported intensity dips
due to the counterbalance between beam self-defocusing and
diffraction. The required negative nonlinearity of the medium
causes an inevitable reduction of the beam intensity. Losses,
saturation and high transverse dimensionality result in
nonintegrable model equations. Despite certain adiabatic
relaxation characteristics, the solitary solutions of these
equations have a large number of characteristics \cite{refc2} in
common with the soliton solution of the one-dimensional NLSE and
are widely denoted by the term ``dark soliton''. DSSs are
generated as dark stripes \cite{refc3,Luther-Davies}, whereas the
only known truly two-dimensional optical DSS is the optical vortex
soliton (OVS) \cite{refc4}. Characteristic of the phase portraits
of these beams are a 1D $\pi $-phase jump and an on-axis $2\pi $
helical phase ramp, also denoted as edge-phase and screw-phase
dislocations, respectively.

Optical ring dark solitary waves (RDSWs)
were first introduced by Kivshar and Yang \cite{refc5}. Their evolution
was studied in the frame of the adiabatic approximation of
perturbation theory. The quasi-one-dimensional treatment allowed
the authors to obtain an expression for the RDSW's transverse
velocity $dR/dz$ as a function of the ring radius and the contrast. In
addition, a linear stability analysis was performed. The authors showed
that in the small-amplitude limit these ring dark waves are
described by the cylindrical Korteweg- de Vries equation, which
has ring soliton solutions \cite{refc5}. Recently,
Frantzeskakis and Malomed derived new
equations of evolution for small-amplitude solitary waves on a
finite background \cite{refc6} by means of a multiscale expansion method
applied to the generalized cylindrical NLSE. Their long-wave solitary solutions
propagate ``on top'' of the continuous-wave solution (i.e. on the
background) as a dark perturbation in Kerr defocusing media.
Depending on the background beam intensity, they can appear both
dark and anti-dark for Kerr-like saturable defocusing
nonlinearity.

The RDSWs may turn out to be of practical interest because of their ability to
induce waveguides in which multiple signal beams could be
guided parallel \cite{Dreischuh}. This requires detailed
analysis of the dynamics of these solitary structures. The first
experimental generation of optical RDSWs \cite{refc7} was conducted
with pure amplitude modulation in front of the nonlinear
medium (reflective micrometer-sized dots illuminated by the
background beam). Under such conditions, pairs of phases oppositely
shifted by about $\pi/2 $ were measured in each diametrical
slice of the RDSWs \cite{refc8}. In a subsequent experiment
\cite{refc9}, RDSWs were generated  from odd initial conditions by
using binary computer-generated holograms (CGHs) \cite{refc10}.
Inside a circle the phase was shifted by $\pi $ vs. the outlying
area by shifting the photolithographically produced interference
lines by half a grating period. In both experiments the phase
profiles were measured by using the multiple-frame interferometric
technique for optical wavefront reconstruction \cite{refc11}. It
was found that along the NLM the transverse velocity of the RDSWs
born from even initial conditions is higher than that of
the odd ones. Numerical simulations have shown that, besides the
ring radius \cite{refc5}, the transverse velocity of the RDSW can be
effectively controlled by initial phase modulation of the
background beam (inside, outside or on both sides of the phase
dislocation) or by nonlinear interaction with a second coaxial
dark formation (RDSW or OVS) \cite{refc12}. In saturable media, the
RDSW's transverse velocity is lower than that in Kerr NLM.
At high saturation levels, in formal contradiction to but in
actual agreement with the one-dimensional case, decay of the dark
rings into a loop of optical vortices of alternate helicities
should be expected \cite{refc13}.

We have measured experimentally the dynamics of ring dark solitary waves
generated by different methods and compared those results with the 
predictions of the analytical method and numerical data. In particular, 
we analyzed the non-monotonic transverse
velocity of RDSWs vs. background beam intensity in saturable NLM. The RDSWs are
generated in the $+1$ and $-1$ diffraction order of the background
beams. Comparative numerical simulations that account
for the estimated contrast of the RDSWs and medium saturation
are carried out. An intentional reduction of the magnitude of the phase shift 
encoded in the CGHs was found to have a substantial
influence on the dark ring radii and the contrast at the entrance of
the NLM. The differences in the nonlinear evolution of the RDSWs
enable us to verify qualitatively the analytical result of Kivshar
and Yang (see Eqs.~6 and 7 in Ref.~\cite{refc5}).

\section{RDSW transverse velocity vs. intensity}
\label{teil1} The experimental arrangement is shown in Fig.~1a. In
order to obtain a dark ring of a desired radius and a phase jump of
a certain magnitude, a single-line $Ar^+$-laser ($\lambda =488nm$) 
is used to reconstruct the respective CGH. The latter are
produced photolithographically with a grating period of $18 \mu m$ 
and ensure a diffraction efficiency of nearly $10\%$ in first
order. The nonlinear medium is ethylene glycol dyed with DODCI
(Lambdachrome). The background beam, diffracted in first order with
the dark ring nested in, is transmitted through a slit placed about
15 cm behind the CGH and is gently focused on the entrance of the
NLM. After passing the desired nonlinear propagation path length
($0.5 cm$ to $8.5 cm$) the dark beam is partially reflected by a
prism immersed in the liquid and is projected directly on a
co-moving CCD array with a resolution of $13 \mu m$.
Alternatively, the dark beam can be recorded at the exit of the
NLM. For further evaluation the images are stored by an image-
capturing card.

The data presented in this section refer to an absorption of
$\alpha=0.107 cm^{-1}$ at $\lambda =488 nm$. In a calibration
measurement we generated a 1D DSS by using a CGH of the
respective type. The soliton constant $Ia^2$ (i.e. the product of
the background-beam intensity $I$ and the square of the dark beam
width $a$ measured at the $1/e^2$-level) is found to reach its
asymptotically constant value for input powers $P_{\rm
sol}^{1D}\approx 90mW$ (Fig.~1b, dashed curve). Since the
saturation of the (thermal) nonlinearity is expected to have
important consequences on the RDSW's nonlinear dynamics \cite{refc14}, 
we realized a background-beam self-bending scheme similar to that used in
\cite{refc15}. The required asymmetry is introduced by
tilting the prism immersed in the NLM. This results
in different propagation path lengths for different parts of
the beam. The strength of the self-bending effect (Fig.~1b, solid
curve) measured in the near field gives an estimated saturation
power $P_{\rm sat}\approx 35 mW$. The choice of a saturation model
for absorptive nonlocal nonlinearity is not trivial
\cite{refc16}. Here we use a sigmoidal fit of the type 
$f(I)\sim I/(1+I/I_{sat})^3$.

\begin{figure}
\centerline{\epsfig{file=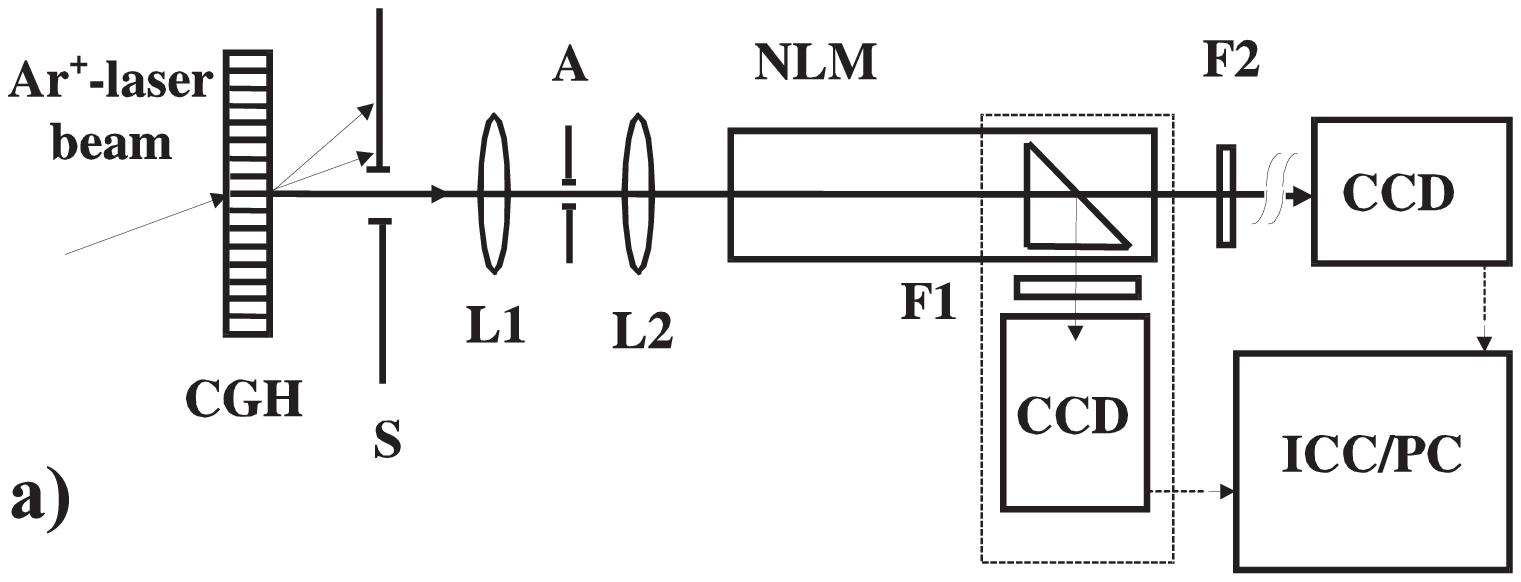,width=3in,clip=} }
\centerline{\epsfig{file=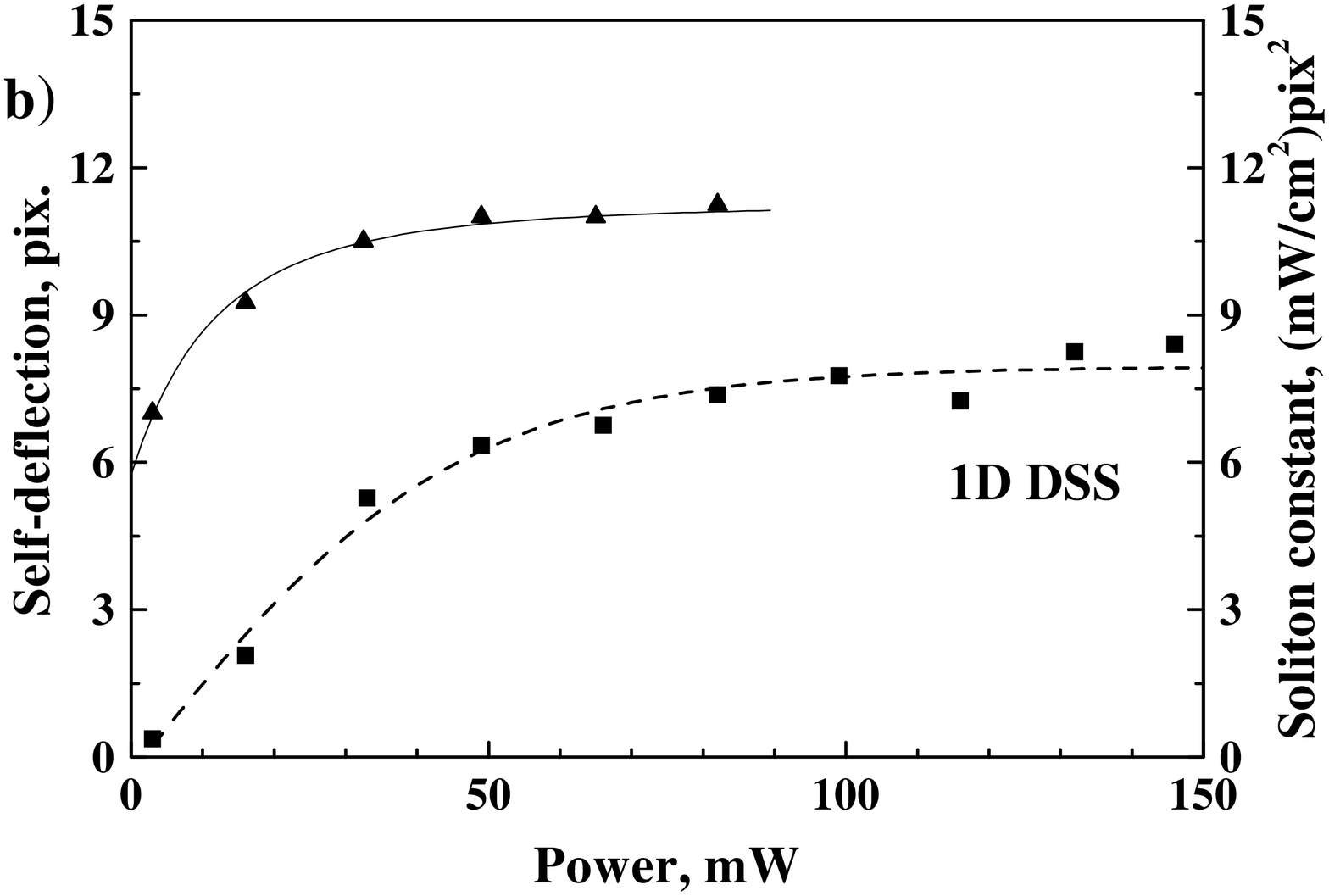,width=3.2in,clip=} }
\caption{{\bf a)} Experimental setup: (CGH - computer-generated
hologram, S - slit, A - aperture, L1 and L2 - AR-coated lenses
($f=80 mm$), NLM - nonlinear medium (ethylene glycol dyed with
DODCI), F1, F2 - filter sets, CCD - charge-coupled device camera
with $13 \mu m$ resolution, ICC/PC - personal computer equipped
with an image-capturing card).
         {\bf b)} Power dependences of the background-beam
self-deflection (triangles) and of the quantity $Ia^2$ (squares)
indicating $P_{\rm sat}\approx 35mW$ and $P_{\rm sol}^{1D}\approx
90mW$. Solid (dashed) curve - sigmoidal fits.}
\label{p8fig1}
\end{figure}

When a RDSW is generated by pure amplitude modulation of the
background beam at the entrance of the NLM, its transverse
velocity increases monotonically with the intensity (up to
$2.5P_{\rm sol}^{1D}$; see Fig.~7 in Ref.~\cite{refc7}). By using the
CGH of a black ring (with a phase jump of $\pi$), we
observe a non-monotonic change of the ring radius vs. background
beam power for a nonlinear propagation length of $z=8.5 cm$
(Fig.~2a). The corresponding data for the contrast of the RDSW
$A^2=(I_{0}-I_{min})/I_0$ show reversed non-monotonic behaviour
(Fig.~2b, solid curve). Generally, this is consistent with
physical intuition and the theory \cite{refc14} that the smaller
the ring, the darker the solitary wave and the higher the total
phase shift. As will be discussed later (see Fig.~4, the region on the
left-hand side of the vertical dashed line), the dark rings initially
generated with a $\pi$-phase shift (``black'') become wider by
diffraction and thus arrive as ``dark-grey'' at the entrance of the
NLM. In our case the contrast is lowered to $A^2=0.88$ (see Fig.~2b). 
We measured the change of the RDSW's radius at $z=8.5 cm$ as a function 
of the laser power. An evident
minimum is observed for a power slightly below $P_{sol}^{1D}$. For
this value the ring contrast is highest. Further insight can be gained by
evaluating the analytical results for the phase of the
solitary wave in the models of competing cubic-quintic and
threshold nonlinearities (see Eqs.~6-9 in Ref.~\cite{refc14}). With
the data for $A^2$, both models show well-pronounced minima in the
RDSW transverse velocity for powers slightly lower than $P_{\rm
sol}^{1D}$ (Fig.~2, dashed curve). It should be noted that the
experimental data in Fig.~2 were obtained by compensating losses during the
propagation with an appropriate change of the input power. However, even for
fixed input power the RDSW changes
its transverse velocity along the medium. In that sense Fig.~2a presents the
effect of the enhancement of the phase shift and the contrast of the RDSW
by saturating the nonlinearity in an integral form. This result, however, does
not explain why in our previous experiments only a monotonic increase
of the RDSW radii was observed \cite{refc7}.

\begin{figure}
\centerline{\epsfig{file=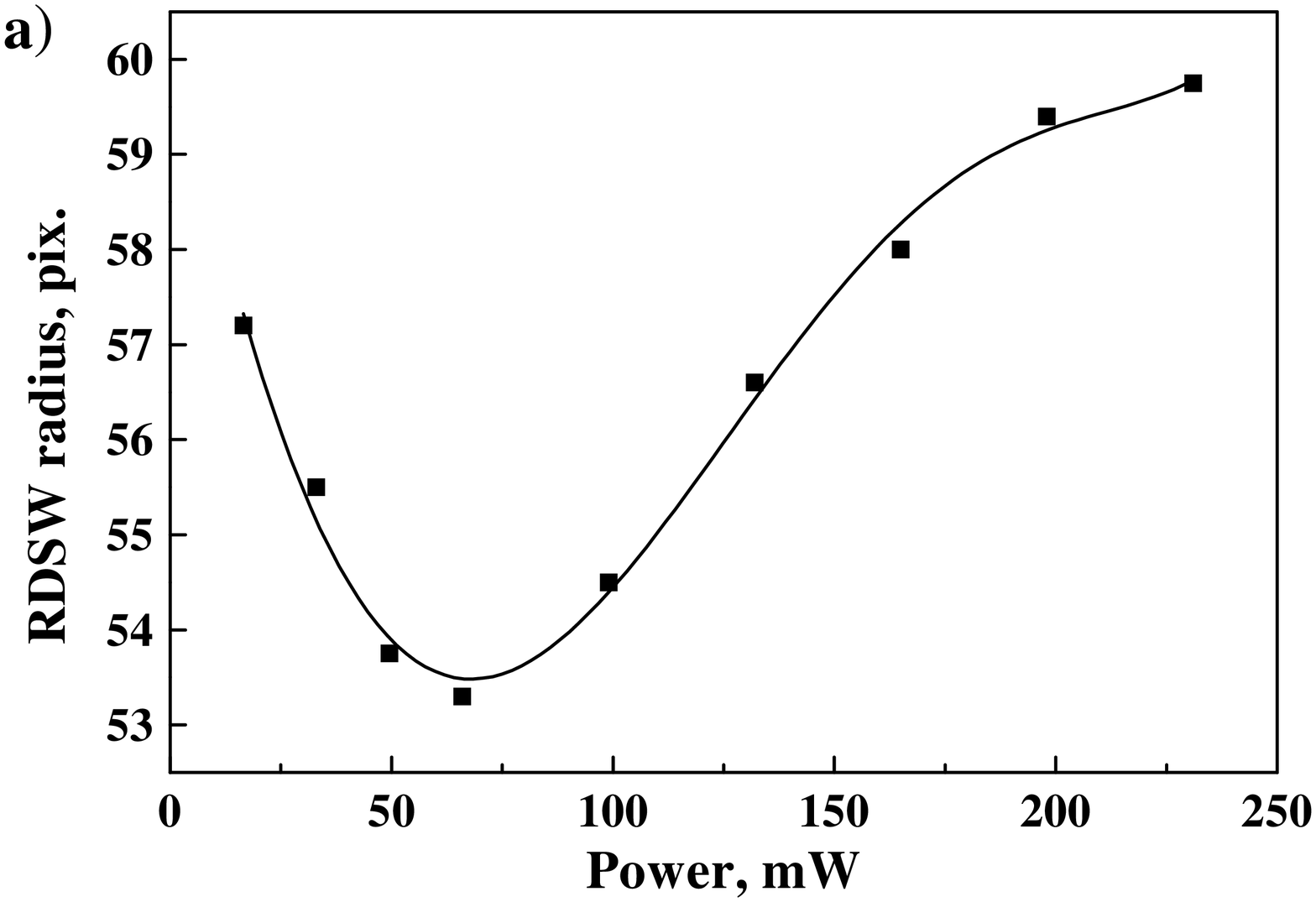,width=2.8in,clip=} }
\centerline{\epsfig{file=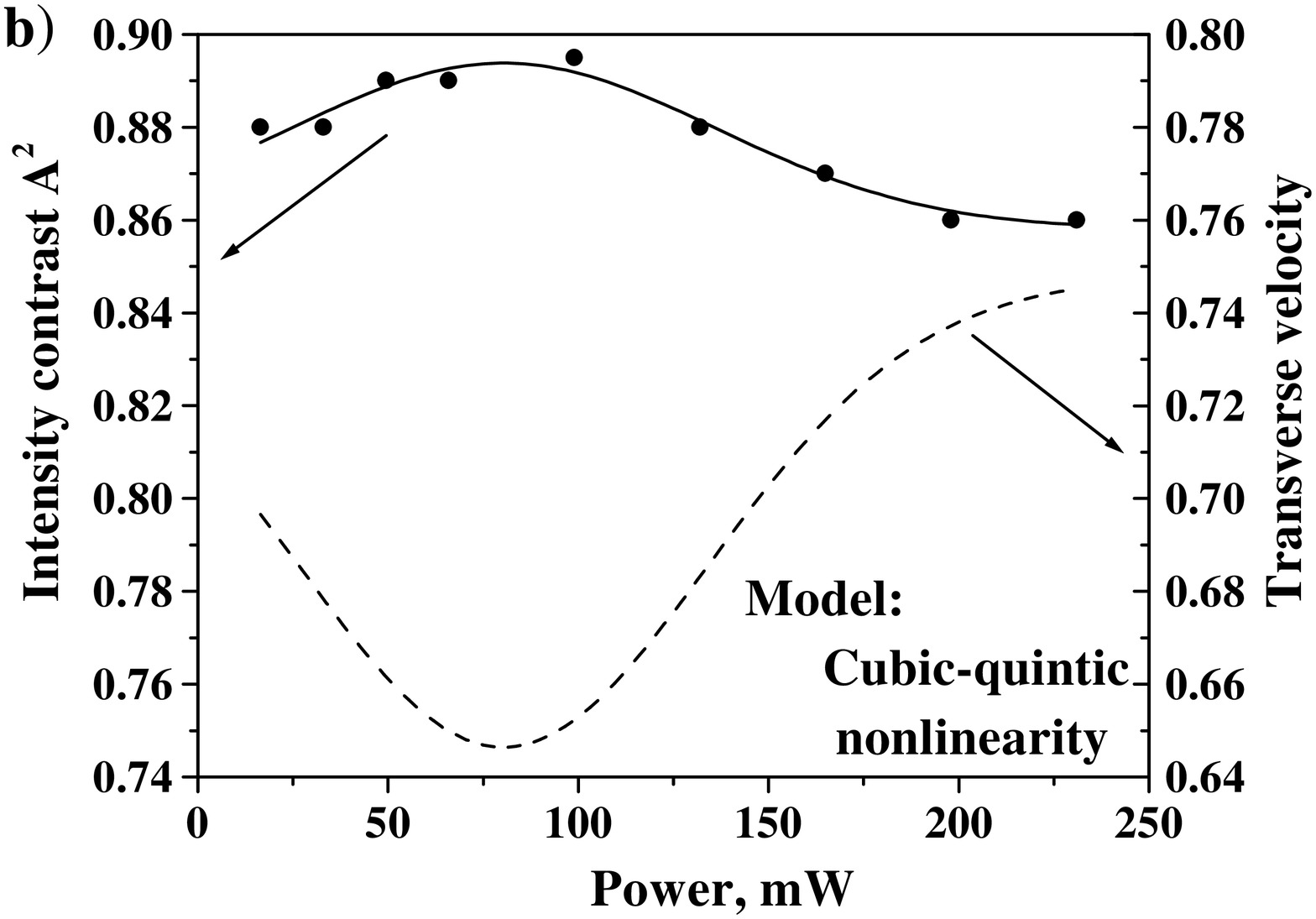,width=2.8in,clip=} }
\caption{{\bf a)} Power dependence of the RDSW radius after a
nonlinear propagation path length $z=8.5 cm$.
         {\bf b)} Power dependence of the RDSW contrast (dots) and estimated
transverse velocity (dashed curve) for saturable cubic-quintic nonlinearity.
See text for details.}
\label{p8fig2}
\end{figure}

In order to explain this discrepancy we look at the evolution of
the RDSW before it enters the NLM. Numerical simulations are
carried out for the evolution of ``grey'' solitary waves with
initial radii $R_0=R(z=0)$ twice as large as the ring width
$r_0=r(z=0)$. The numerical procedure used to solve the 2D NLSE is
based on the beam propagation method over a $1024\times 1024$
grid. In agreement with our measurement, initially ``black'' rings
generated by CGHs become ``grey'' after passing four Rayleigh
diffraction lengths $L_{Diff}=kr_0^2$ and reach the estimated
contrast $A^2=0.88$. The evolution of the RDSW's radius for
different intensities is calculated for up to $10$ nonlinear
lengths ($L_{NL}=(k|n_2|I_0)^{-1}$, $k$ is the wavenumber, and $|n_2|I_0$
the nonlinear refractive index correction). The behaviour of the RDSW
radius in the NLM can be summarized as follows:\\
   i) For a fixed input intensity in the range of $(0.2-2.2)I_{\rm sol}^{1D}$
all RDSWs have their minimal ring radius for propagation distances
between $1.5L_{NL}$ and $3L_{NL}$.\\
   ii) For nonlinear propagation distances shorter than $2L_{NL}$ the grey rings
monotonically reduce their radii for increasing intensity (Fig.~3, 
lowest box).\\
   iii) For long propagation distances ($>6L_{NL}$)
this tendency is reversed (Fig.~3, upper box).\\
   iv) In between, the radius of the RDSW as a function of the intensity
shows a characteristic minimum around $4.5L_{NL}$ (Fig.~3,
middle box). This propagation distance is in
reasonable agreement with the NLM length of $8.5 cm$ corresponding
to $4L_{NL}$, for which Fig.~2 is obtained.

\begin{figure}
\centerline{\epsfig{file=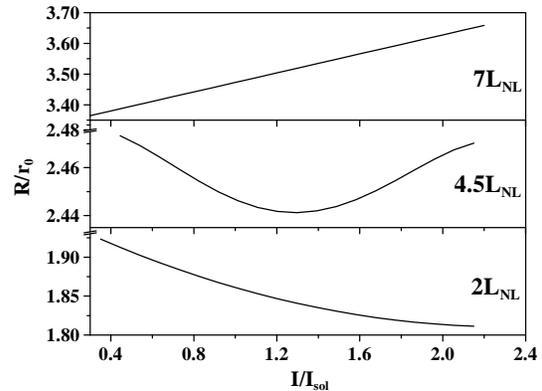,width=3.0in,clip=} }
\caption{RDSW radius vs. normalized background-beam intensity at
three characteristic nonlinear propagation distances. Model
parameters: $R_0/r_0=2$, $A^2=0.88$, $s=0.4$.}
\label{p8fig3}
\end{figure}

Generally, we found qualitative confirmation of the existence of
a minimal radius $R_{min}$ of the RDSW, which depends on the initial
contrast $A^2$ \cite{refc5,refc14}. The evolution of the contrast
itself is strongly influenced by the saturation of the
nonlinearity. In view of the above it is interesting to recall the
observation which we made when modeling the decay of multiple-charged
OVSs under radial perturbations \cite{refc17}: The small-amplitude
even-phase initial conditions for perturbation and the
saturation of the nonlinearity resulted in a negative transverse
velocity of the rings. Their
contrast was found to increase with decreasing radius and the radial phase
variation reached a maximum of $0.75\pi$ at $5L_{NL}$. The initially
decaying and still strongly overlapping optical vortices partially
recovered their multiple-charged state (Fig.~7 in Ref.~\cite{refc17}).
Periodic sign changes in the transverse velocity of the dark rings
in the propagation direction were observed in the simulations up
to $10L_{NL}$ with a period of approximately $2L_{NL}$.

\section{RDSW transverse dynamics vs. initial phase shift}
\label{teil2}
It should be pointed out that initially ``black'' RDSWs with high
transverse dynamics are not likely to be present in a real
experiment, at least when generated  by holograms. The reason lies
in the necessity to select one of the diffracted beams, which
requires a certain free-space propagation. After some diffraction
is ``accumulated'', the dark rings with small radii spread out
substantially and their contrast decreases. Dark rings with large
radii are less affected and they evolve more slowly inside the NLM.
Figure 4 is intended to visualize this before and inside the NLM.
The coordinate $z=0$ is fixed on the nonlinear interface. Solid
and dashed lines correspond to Kerr and saturable Kerr
nonlinearity ($s=I/I_{sat}=0.4$). The simulation presented is
carried out for $I=I_{\rm sol}^{1D}$ and for an initial phase
jump of $\pi $, which will be flattened in the course of propagation.
Generally, the saturation of the nonlinearity leads to a reduction of the
dynamics of the RDSW. For our further considerations it is important to
note that there is no difference whether the phase inside the ring
$\Phi_{in}$ is bigger or lower than the phase $\Phi_{out}$
outside the ring, provided $\Delta\Phi=|\Phi_{in} -
\Phi_{out}|=\pi$. Equivalently, there should be no difference in
the transverse dynamics of the RDSWs generated by the $+1$st and
$-1$st diffraction order beam when a phase jump of $\pi $ is encoded
in the CGHs. This was carefully proved by numerical simulations
and experimentally.

\begin{figure}
\centerline{\epsfig{file=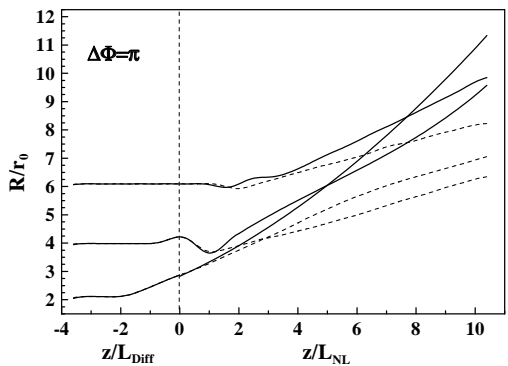,width=3.0in,clip=} }
\caption{Diffraction-governed evolution of odd RDSWs with initial $\pi $-phase
jumps over 4 Rayleigh diffraction lengths followed by nonlinear propagation in 
a Kerr (solid curves) and saturable Kerr-like medium (dashed curves; $s=0.4$). 
$I=I_{\rm sol}^{1D}$. Vertical dashed line - the nonlinear interface.}
\label{p8fig4}
\end{figure}

\subsection{Linear evolution for $|\Delta\Phi|<\pi$}
\label{teil2.1}
The situation changes when we produce dark ring waves encoded on
CGHs with phase shifts smaller than $\pi$. Unfortunately, in
binary CGHs only $\pi /N$ steps are possible, where one grating
period consists of $2N$ elementary stripes. In our case $N=3$ and
the grey rings are designed for $|\Delta\Phi|=2\pi /3$. The beam
of a single-mode He-Ne laser is expanded 3.5 times and reproduces
the respective CGH. By means of a  lens ($f=4 cm$) the beam
transmitted directly and the $\pm 1$st order diffracted beams are
imaged on a screen located about $100 cm$ behind.  By varying the
lens position we are able to follow the dark ring evolution.
Figure 5a shows the images of the still overlapping first-order
diffracted beams, which interfere with the zeroth order one. It
can be seen clearly that the rings with $|\Delta\Phi|=2\pi /3$
have smaller (larger) radius than each of
the dark rings with an initial phase shift $|\Delta\Phi|=\pi $.

\begin{figure}
\centerline{\epsfig{file=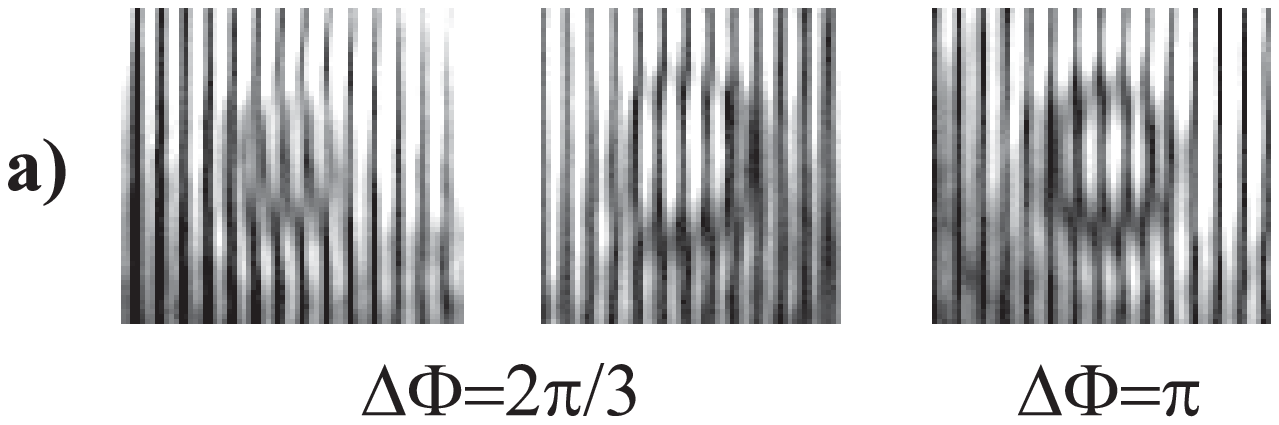,width=3in,clip=} }
\centerline{\epsfig{file=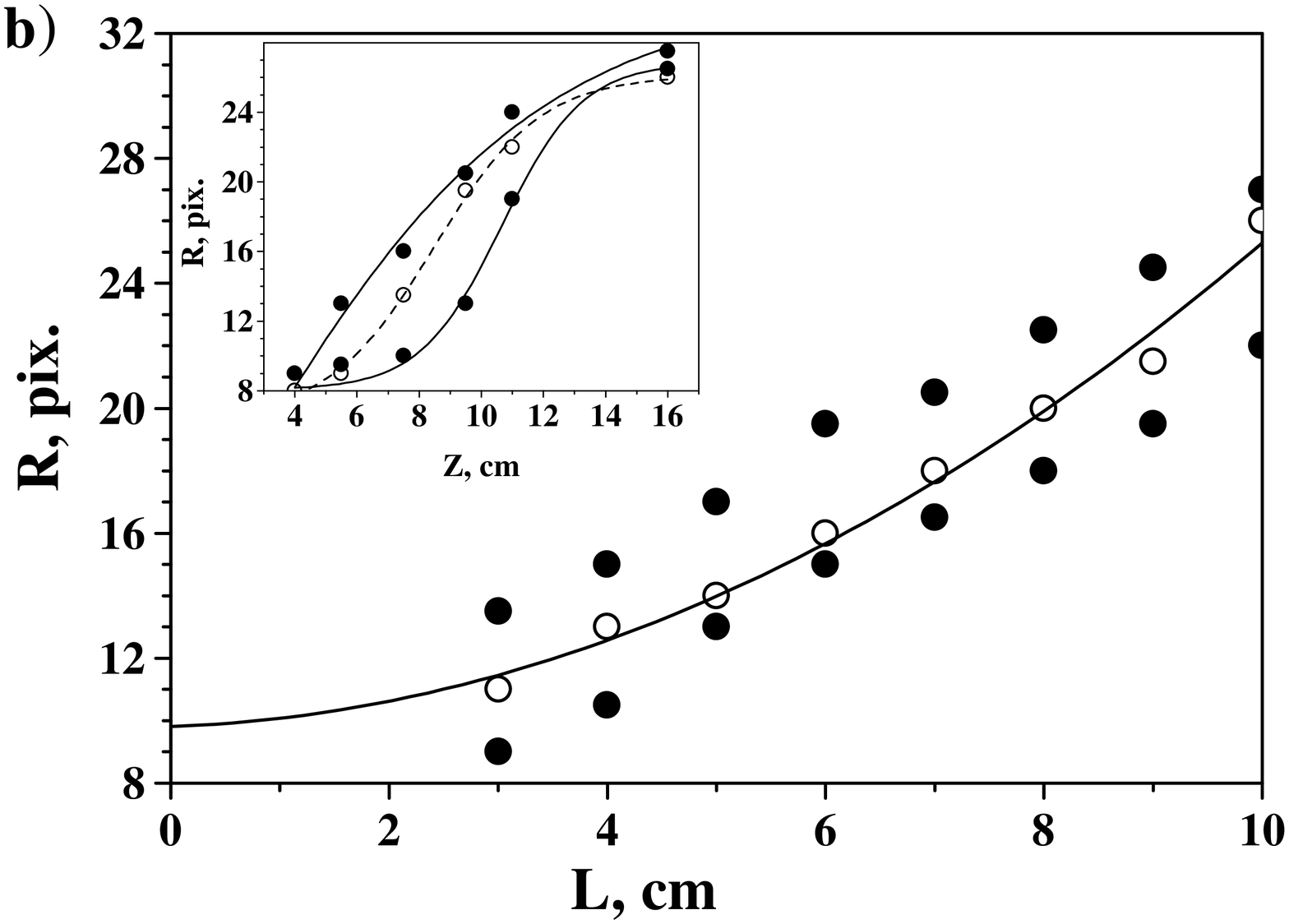,width=3.0in,clip=} }
\caption{Linear propagation of ring dark beams:
   {\bf a)} Interference patterns indicating different transverse velocities of
free-propagating, initially grey ($|\Delta\Phi|=2\pi /3$) and black
($|\Delta\Phi|=\pi $) ring dark beams at $L=5.5 cm$.
   {\bf b)} Host graph: ring radii vs. CGH-to-CCD-array distance L.
Inset: the same vs. lens-to-CGH-distance. Dots - $\pm 1$st diffraction order
beams, blank circles - initially black dark rings. See text for details.}
\label{p8fig5}
\end{figure}

The ring radii observed for lens positions from $4 cm$ to $16 cm$
are shown in the inset of Fig.~5b. In order to get better
spatial resolution, the measurement was repeated with a CCD array
illuminated directly after the CGH (Fig.~5b). Blank circles denote the
dark ring radii for $|\Delta\Phi|=\pi $, the solid circles refer to
the ring with encoded phase jump $|\Delta\Phi|=2\pi /3$. Note
that in the two graphs of Fig.~5b the ring radius $R$ is denoted
in CCD-camera pixels, but these units are not directly comparable because of
the different focusing conditions. On the CGHs the ring radius is encoded
in 20 pix. (i.e. $R_{0}=60 \mu m$). It is easy to understand the
observed differences: The diffraction tends to broaden the dark
rings and flatten the phase jumps. As a result, for
$\Phi_{in}<\Phi_{out}$ an effective concave phasefront is formed
which ``focuses'' the dark ring. In the other diffraction order
$\Phi_{out}$ appears to be bigger than $\Phi_{in}$. The
phase change induced by the diffraction results in an effective convex
phasefront that increases the dark ring radius. For
$|\Delta\Phi|=\pi $, the diffraction influences both first-order
beams in the same way, but more weakly.

\subsection{Nonlinear evolution of initially grey RDSWs}
\label{teil2.2}
Although the nonlinear regime was analytically analyzed by Kivshar
and Yang \cite{refc5}, to the best of our knowledge it has not
been investigated experimentally. The basic results (Eqs.~5-7 in
Ref~\cite{refc5}) can be summarized as follows:\\
   i) For each RDSW there is a minimum ring radius $R_{min}$ for
which the RDSW has its highest contrast.\\
   ii) Depending on the initial value of the phase jump, the RDSW
either collapses along the NLM to reach $R_{min}$ and
diverge monotonically thereafter, or it diverges monotonically from the
beginning.\\ 
\begin{figure}
\centerline{\epsfig{file=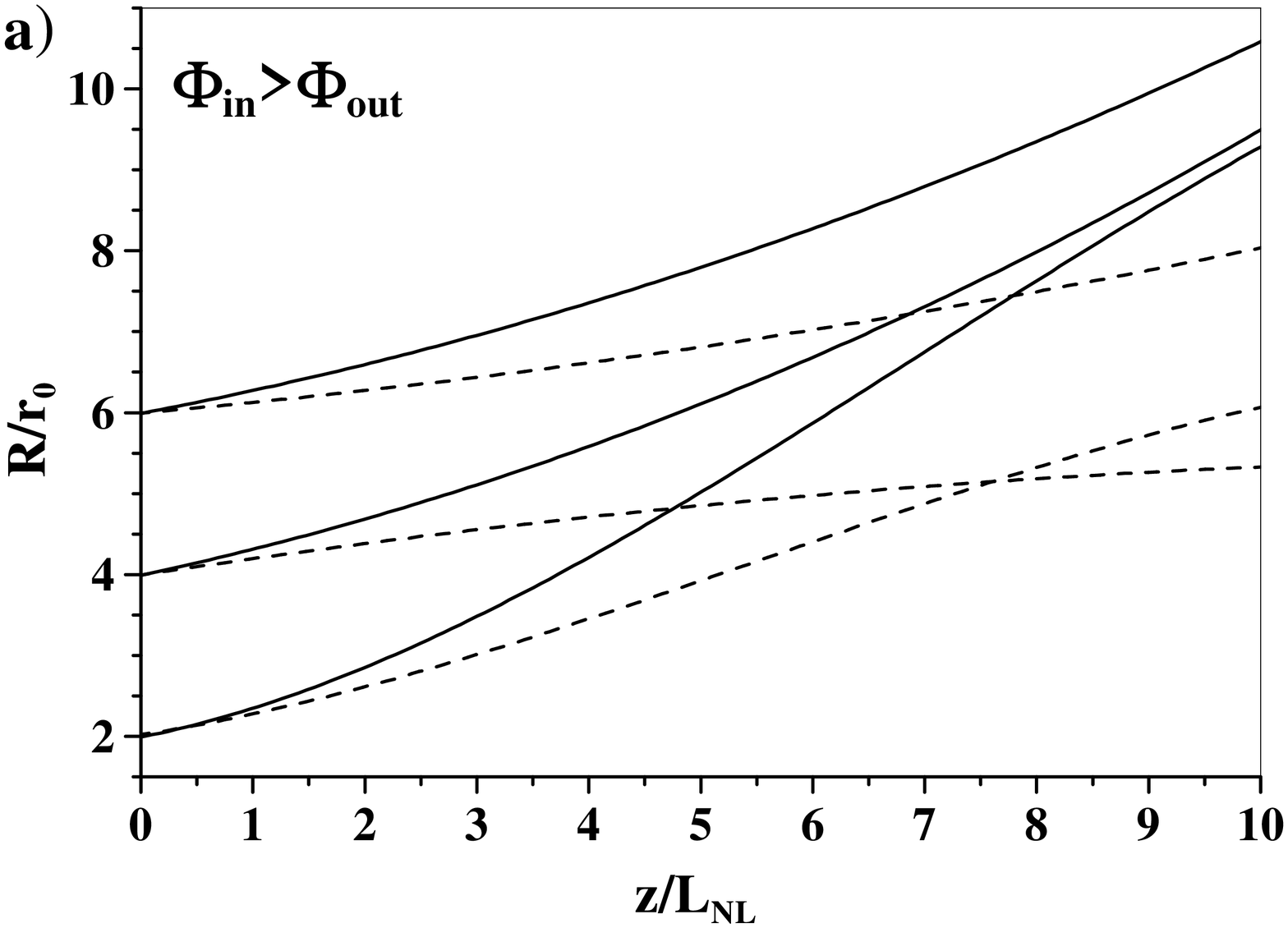,width=2.85in,clip=} }
\centerline{\epsfig{file=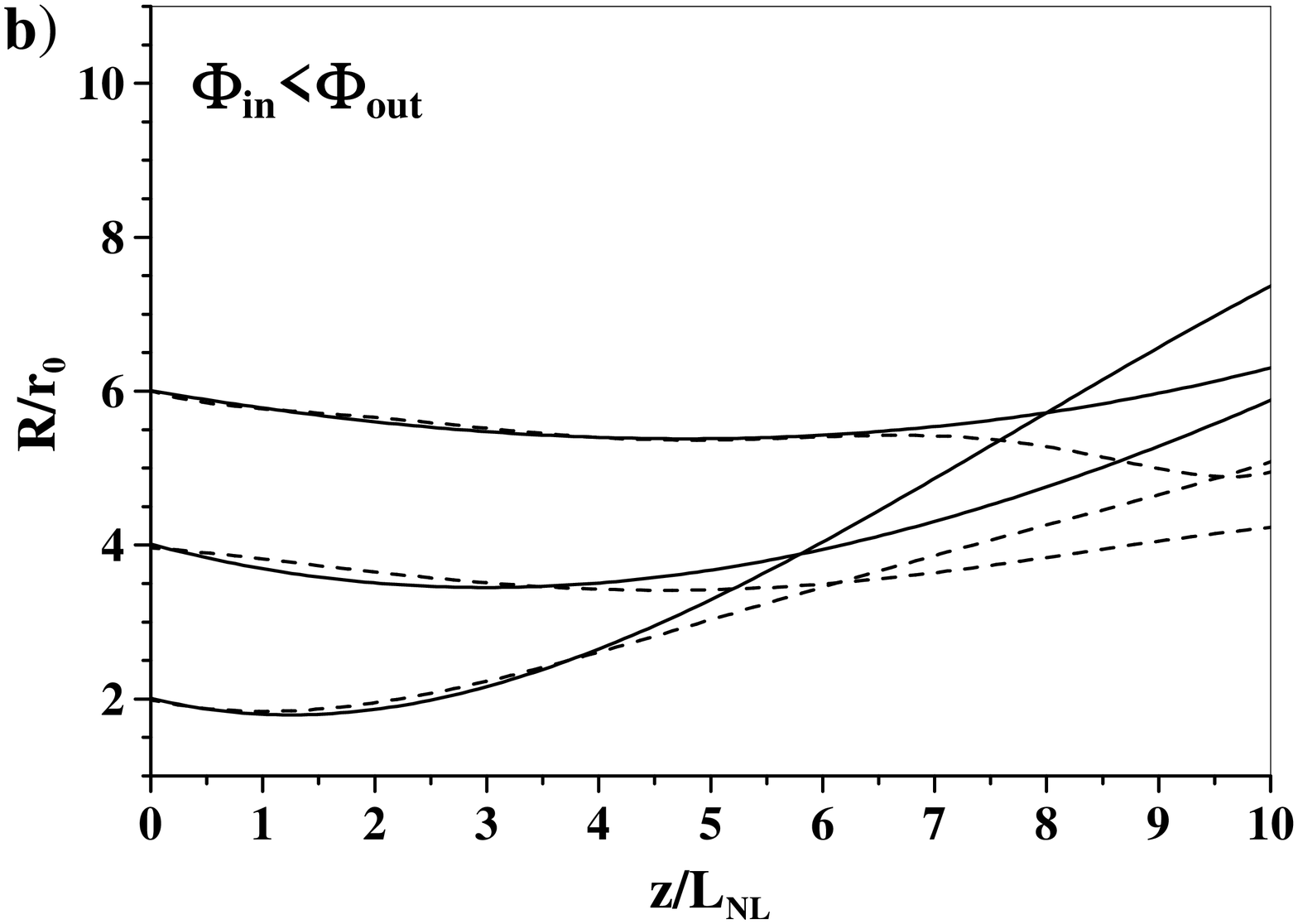,width=2.85in,clip=} }
\caption{Nonlinear evolution of initially grey RDSWs in a Kerr
(solid curves) and saturable Kerr-like medium (dashed; $s=0.4$) for
positive {\bf (a)} and negative {\bf (b)} phase shift
$\Delta\Phi=\Phi_{in} - \Phi_{out}$. The model parameters
correspond to $A^2=0.88$ (i.e. to $|\Delta\Phi|=0.837\pi$) and
$I=I_{\rm sol}^{1D}$.}
\label{p8fig6}
\end{figure}

Figure 6 is intended to clarify these conclusions. We
solved the NLSE numerically by keeping the following parameters
close to the experimental values: RDSW radii ($(R/r)|_{z=0} =2,4,$ and $6$), 
background-beam intensity ($I=I_{\rm sol}^{1D}$),
initial contrast $A^2=0.88$, phase jump
$|\Delta\Phi|=0.837\pi$, and saturation of the nonlinearity $s=0.4$. In
Fig.~6a the corresponding results for the case of a higher phase
inside the rings ($\Phi_{in}>\Phi_{out}$) are presented. In this
case, the RDSWs start diverging immediately after entering
the NLM. The data shown in Fig.~6b refer to the case of a lower
phase inside the dark ring and higher phase outside
($\Phi_{in}<\Phi_{out}$). The effective concave wavefront forces
the RDSWs to collapse to a minimum ring diameter. Thereafter they
start to broaden faster. The main tendency, that the broader the
RDSW initially is, the longer the nonlinear propagation needed to reach
$R_{min}$ will be, is well expressed. The other general tendency, that the
smaller the RDSW diameter is, the higher its transverse dynamics is, can also
be clearly seen.

Both tendencies and the influence of the saturation seem to be
understood theoretically, but still need experimental confirmation. We
performed additional measurements with the experimental setup
shown in Fig.~1, in which the telescope is replaced by a single lens ($f=12cm$)
in order to focus the beams near the entrance of the NLM. Moderate and
strong saturation $s$ can be achieved by raising the absorption
$\alpha$ to $0.2 cm^{-1}$ ($P_{\rm sol}^{1D}=25 mW; P_{\rm sat}=40
mW; s=0.6$) and $0.4 cm^{-1}$ ($P_{\rm sol}^{1D}=20 mW; P_{\rm
sat}=14(\pm 5) mW; s\approx 1.4$). In order to keep
the conditions for both first-order beams as similar as possible,
the beam transmitted directly was blocked at the entrance of the
NLM and both $\pm 1$st order beams were recorded simultaneously.
Unfortunately, due to the interaction of the two background beams,
the interaction-free nonlinear propagation path length is limited.
The results of the measurements are shown in Fig.~7a for moderate
and high saturation (host graph and inset, respectively). Blank
circles denote ``black'' RDSWs radii, solid ones RDSW radii for
$\Delta\Phi=\pm 2\pi /3$. The best results for moderate saturation are
achieved with CGHs containing phase dislocations with radii of $R=30 \mu m$. 
For high saturation, rings with dislocation radii of $60 \mu m$
are used. As seen, after some $5 mm$ of nonlinear propagation the
``grey'' RDSWs reach $R_{min}$ and diverge monotonically afterwards.
The beam of the other diffraction order and the RDSW generated
for $|\Delta\Phi|=\pi$ start diverging from the beginning. The
interpolated curves in the figure are intended to guide the eye.
The interpolated curves for $z<2 mm$ and $z>17 mm$ are not shown
for an obvious physical reason$:$ As mentioned, the grey RDSWs do
not enter the NLM with equal radii and equal (absolute) phase
shifts. For larger propagation distances, the saturation-like
behaviour is caused mainly by interaction with the second
co-propagating beam. The reduced transverse velocities of the RDSWs
for higher saturation are also strongly pronounced. The grey-scale
images in Fig.~7b show the RDSWs at $z=2 mm$ and $8 mm$ inside the NLM.

\begin{figure}
\centerline{\epsfig{file=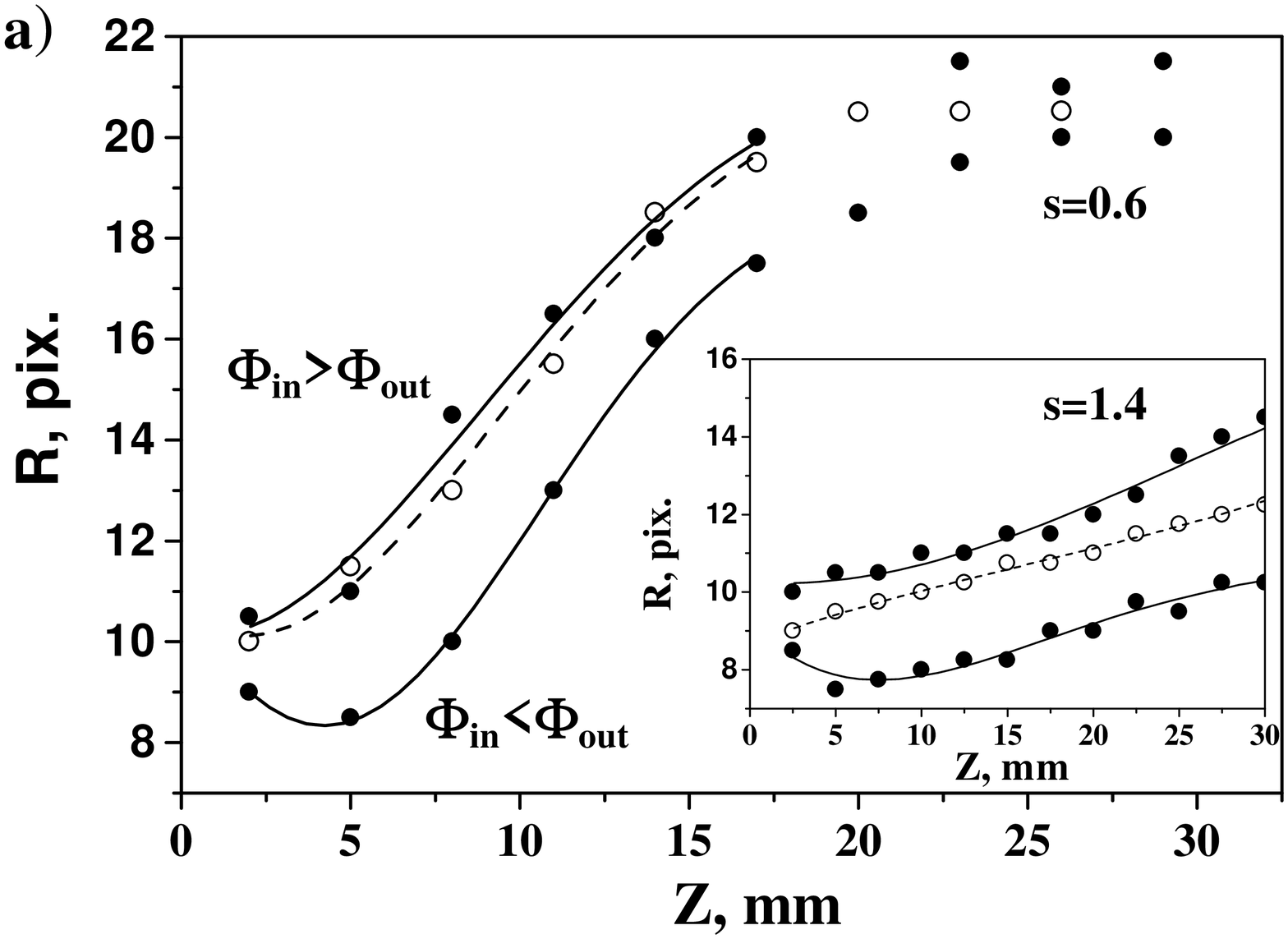,width=3.2in,clip=} }
\centerline{\epsfig{file=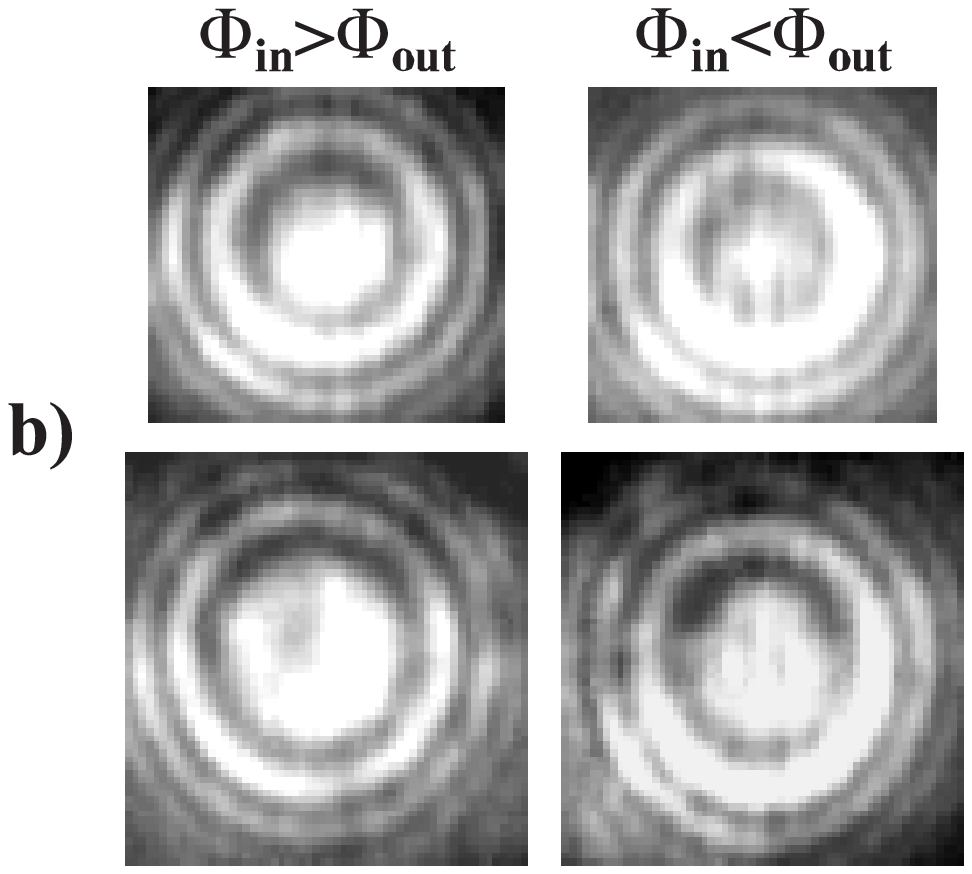,width=2.0in,clip=} }
\caption{Nonlinear propagation of RDSWs:
         {\bf a)} Evolution of the RDSW radii along the nonlinear medium
at moderate (host graph, $s=0.6$) and high saturation (inset,
$s\approx 1.4$). Solid and blank circles - initially grey
($|\Delta\Phi|=2\pi /3$) and black ($|\Delta\Phi|=\pi $) RDSWs, respectively.
         {\bf b)} Grey-scale images of RDSWs at $z=2 mm$ (upper row) and
$8 mm$ (lower row) in a moderately saturated NLM.}
\label{p8fig7}
\end{figure}

\begin{figure}
\centerline{\epsfig{file=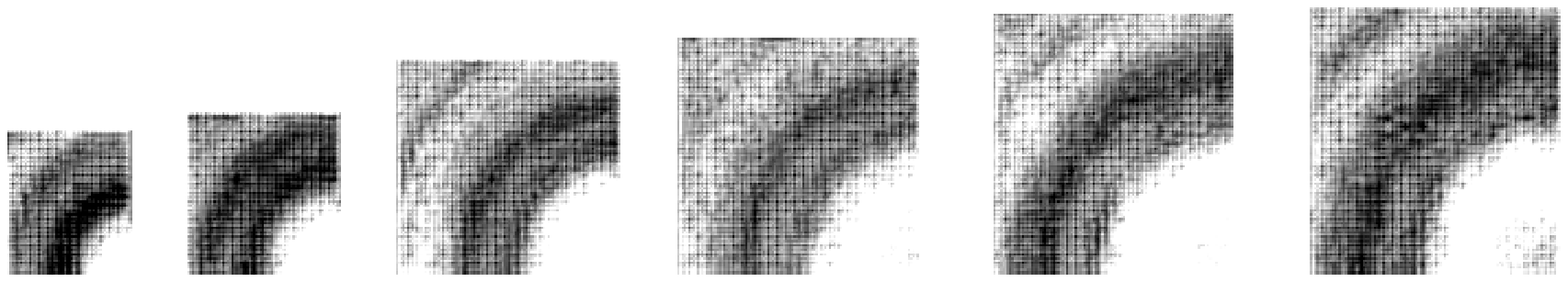,width=3.5in,clip=} }
\caption{ Gray-scale images showing the development of an internal
substructure of a RDSW ($R_0=90 \mu m$) at moderate saturation.
Left to right: $z=2, 5, 8, 11, 14,$ and $17  mm$.}
\label{p8fig8}
\end{figure}

By evaluating the data corresponding to the largest RDSWs in our
experiment ($R=90 \mu m$) we observed internal ring splitting
for moderate saturation (Fig.~8). Despite their large radii,
these RDSWs have non-zero transverse velocity.
They broaden and flatten due to the saturation (see Fig.~8,
frame at $z=2 mm$) while evolving into grey rings. It seems that such a grey
ring can serve as a background on which two coaxial rings can form,
subsequently narrow
and thus get a high modulation depth for increased propagation distance
(Fig.~8, frames at $z=5, 8, 11, 14,$ and $17 mm$). The mechanism
for their formation includes effects from increasing a phase shift and
decreasing a transverse velocity due to the saturation of the nonlinearity
\cite{refc14}, as well as repulsion between opposite quasi-2D
phase dislocations. The relation of these coaxial rings to the
small-amplitude dark solitary waves described by Frantzeskakis and
Malomed \cite{refc6} should be clarified.

\section{Conclusion}
It is worth noting that it is difficult (if possible at all) to
ensure perfect odd initial conditions (radial phase jumps equal to
$\pi$ and contrast equal to unity) for the generation of RDSWs in
cubic nonlinear media. When such a phase jump is encoded in a CGH,
there is no difference between the evolution of the two beams in
first diffraction order. This changes gradually for smaller phase
jumps. In qualitative agreement with the theory \cite{refc5} we
presented evidence of the monotonic increase of the dark ring
radius for $\Phi_{in}>\Phi_{out}$ and the corresponding
non-monotonicity in the opposite case. For grey initial conditions
and saturation of the nonlinearity we found characteristic regions
for the dependence of the RDSW radius on the background-beam
intensity. The radius can increase, decrease monotonically or go
through a minimum. The results presented are in qualitative
agreement with the analytical ones regarding the phase of the solitary wave
in the presence of saturation \cite{refc14}.

\acknowledgments The authors are pleased to acknowledge highly useful
discussions with Prof. Yu. Kivshar and his suggestions for improving the
manuscript. A. D. would like to thank
the Alexander von Humboldt Foundation for the award of a fellowship and the
opportunity to work in the stimulating atmosphere of
Max-Planck-Institut f\"ur Quantenoptik (Garching, Germany). This
work was also supported by the National Science Foundation of
Bulgaria and by the Science Fund of Sofia University.

\end{document}